\documentclass[11pt]{article}
\usepackage{graphics}
\def \perc{\small{\%}}
\def \lp{\scriptscriptstyle +}

\def\sss{\rm_S}
\def\phrac#1/#2{\leavevmode\kern.1em\raise.5ex\hbox{\the\scriptfont0 #1}
\kern-.1em/\kern-.15em\lower.25ex\hbox{\the\scriptfont0 #2}}
\begin{document}
\parskip 3pt plus 1pt
\baselineskip=12 pt plus 2pt minus 1pt
\null
\def\moriond{\vtop{\hbox{UMS/HEP/92-020}
                   \hbox{XXVII$^{th}$ Rencontre de Moriond}
                   \hbox{Electroweak Interactions and Unified Theories}
                   \hbox{Les Arcs, France (15--22 March 1992) 417--422}}}
\rightline{\moriond}
\vskip 16pt
\centerline {\bf CHARM \thinspace PHYSICS \thinspace AT \thinspace
FERMILAB \thinspace E791}
\vskip 8pt
\noindent
D.J. Summers,$^6$
E.M. Aitala,$^6$
F.M.L. Almeida,$^9$
S.\ Amato,$^1$
J.C. Anjos,$^1$
J.A.~Appel,$^4$
D.~Ashery,$^{10}$
J.~Astorga,$^{12}$
S.~Banerjee,$^4$
I.~Bediaga,$^1$
G.~Blaylock,$^2$
S.~B.~Bracker,$^{11}$
P.R.~Burchat,$^2$
R.~Burnstein,$^5$
T.~Carter,$^4$
I.~Costa,$^1$
L.M.~Cremaldi,$^6$
K.~Denisenko,$^4$
C.~Darling,$^{14}$
P.~Gagnon,$^2$
S.~Gerzon,$^{10}$
K.~Gounder,$^6$ 
D.~Granite,$^7$
M.~Halling,$^4$
C.~James,$^4$
P.A.~Kasper,$^5$
S.~Kwan,$^4$
J.~Lichtenstadt,$^{10}$
B.~Lundberg,$^4$
\hbox{S.~May--Tal--Beck,$^{10}$}
J.~R.~T.~de~Mello~Neto,$^1$
R.~Milburn,$^{12}$
J.~de Miranda,$^1$
A.~Napier,$^{12}$
A.~Nguyen,$^7$
A.B.~de~Oliveira,$^3$
K.C.~Peng,$^5$
M.V.~Purohit,$^8$
B.~Quinn,$^6$
S.~Radeztsky,$^{13}$
A.~Rafatian,$^6$
A.J.~Ramalho,$^9$
N.W.~Reay,$^7$
K.~Reibel,$^7$
J.J.~Reidy,$^6$
H.~Rubin,$^5$
A.~Santha,$^3$
A.F.S.~Santoro,$^1$
A.~Schwartz,$^8$
M.~Sheaff,$^{13}$
R.~Sidwell,$^7$
H.~da~Silva~Carvalho,$^9$
J.~Slaughter,$^{14}$
M.D.~Sokoloff,$^3$
M.~H.~G.~Souza,$^1$
N.~Stanton,$^7$
K.~Sugano,$^2$
S.\ Takach,$^{14}$
K.\ Thorne,$^4$
A.~Tripathi,$^7$
D.~Trumer,$^{10}$
J.~Wiener,$^8$
N.\ Witchey,$^7$
E.\ Wolin,$^{14}$
D.\ Yi$^6$

{\footnotesize \baselineskip=7 pt plus 1pt 
$$\vbox {\halign{#\hfil & \quad #\hfil \cr
{$^1$Centro Brasileiro de Pesquisas F{\'\i}sicas, Rio de Janeiro} &
{$^2$Univ.\ of California, Santa Cruz, CA 95064}             \cr
{$^3$University of Cincinnati, Cincinnati, OH 45221}             &
{$^4$Fermilab, Batavia, IL 60510}                                \cr
{$^5$Illinois Institute of Technology, Chicago, IL 60616}        &
{$^6$University of Mississippi, Oxford, MS 38677}                \cr
{$^7$Ohio State University, Columbus, OH 43210}                  &
{$^8$Princeton University, Princeton, NJ 08544}                  \cr
{$^9$Universidade Federal do Rio de Janeiro, Rio de Janeiro}     &
{$^{10}$Tel--Aviv University, Tel--Aviv 69978}                   \cr
{$^{11}$317 Belsize Drive, Toronto, Ontario M4S1M7 \ Canada}       &
{$^{12}$Tufts University, Medford, MA 02155}                     \cr
{$^{13}$University of Wisconsin, Madison, WI 53706}              &
{$^{14}$Yale University, New Haven, CT 06511}                    \cr
}}$$
}

\vbox{
\centerline {Presented by \ D. J. Summers}
\centerline{Dept. of Physics and Astronomy, University of Mississippi}
\centerline{Oxford, Mississippi 38677 \ USA}}
\vspace*{5mm}
\leftline{\hspace{54mm}
\resizebox{40mm}{!}{\includegraphics{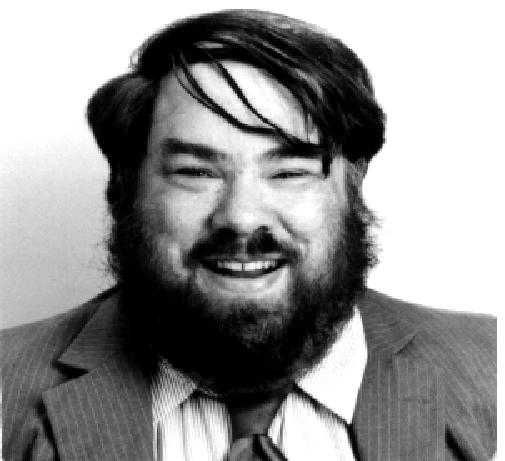}}}
\vfill
\leftline {\bf ABSTRACT}
\noindent
Experiment 791 at Fermilab's Tagged Photon Laboratory has just
accumulated a high statistics charm sample by recording 20 billion
events on 24000 8mm tapes.  A 500 GeV/c $\pi^-$ beam was used with a
fixed target and a magnetic spectrometer which now includes 23 silicon
microstrip planes for vertex reconstruction.  A new data acquisition
system read out 9000 events/sec during the part of the Tevatron cycle
that delivered beam.  Digitization and readout took 50 $\mu$S per
event.
Data was
buffered in eight large FIFO memories to allow continuous event building and
continuous tape writing to a wall of 42 Exabytes at 9.6 MB/sec.
The 50 terabytes of data buffered to tape is now being filtered on RISC
CPUs. Preliminary results show $D^0 \rightarrow K^-\pi^+$ and 
$D^+ \rightarrow K^-\pi^+\pi^+$ decays.  Rarer decays will be pursued. 
\eject
\leftline{\bf INTRODUCTION}
\par
   In the E791 proposal to Fermilab, we presented a plan to collect and
fully reconstruct a minimum of 100,000 mesons and baryons containing a
charm quark. A primary motivation for this experiment is to search for
rare charm decays that probe the standard model. This was the fourth in
a series of charm experiments at the Tagged Photon Laboratory (TPL) with the
same underlying spectrometer: E516[1], E691[2], E769[3], and now E791. 
With the exception of E516, we have used an open, unsophisticated
trigger and have seen large numbers of charm particles by recording and
reconstructing large numbers of events. Mesons and baryons with a charm
quark are characterized by a high mass and a decay length of a few
millimeters to a couple of centimeters. These two characteristics are
difficult, expensive, and risky to trigger on.  A similar problem exists
for a large fraction of beauty decays at hadron colliders.  At the time
of the 791 proposal, three developments made it possible to continue the
conservative TPL philosophy of an open trigger. New fast front ends made
it possible to digitize and read out events in tens of micro--seconds.
Exabyte tape drives and 8mm tapes made it possible to store and handle
large amounts of
data at a reasonable cost.  RISC CPU chips, such as the MIPS R3000 and
its rapidly improving cousins, made it possible to keep the offline
computing budget below a quarter of the total E769 $\rightarrow$ E791
upgrade cost. 
\vskip 10pt
\leftline{\bf BEAM CHOICE, TRIGGER, TARGET DISCS, AND BEAM TRACKING}
\par  

   At the Tagged Photon Lab we now use pions rather than photons.
High energy
photons produce more charm per hadronic interaction than pions, but it
is just not possible to get enough photons. We chose to use
pions rather than protons because pions have a stiffer
gluon distribution for a given momentum. Charm is primarily produced by
gluon--gluon fusion. 
We doubled the E769 beam momentum of 250 to 500 GeV/c for E791 
because this is
estimated to increase the charm cross section by 80\% and to triple the
miniscule beauty cross section [4].  The beam was negative; a
$\pi^+$ beam would have been heavily contaminated with protons. 
\par
   A loose $E_T$ trigger was used to enrich charm and reduce the raw
interaction trigger rate.  Energy was measured in electromagnetic and
hadronic calorimeters [5].  Events were rejected if two beam particles
in coincidence might fake an $E_T$ trigger. The beam rate was
2 MHz.
\par
   The target consists of five parts, one 0.5mm thick platinum disk
followed by four 1.6mm thick diamond disks.  Fig. 1 shows the {\it z}
positions of the disks as determined by primary vertices found by our
Silicon Microvertex Detector (SMD).  The thinness of the disks provides
a strong constraint on the {\it z} position of a primary vertex and the
air gaps in between provide volumes for reconstructing secondary
vertices uncontaminated by secondary interactions. Dense materials
($\rho$ = 21.4 and 3.3 g/cc) were chosen to allow thin targets which
would still cause 
\insert\footins{
$$\vbox {\halign{#\hfil & \ \ #\hfil & \quad #\hfil \cr
{\bf Table 1.}      &                               &                       \cr
Spectrometer Upgrade & E769 $\longrightarrow$       &   E791                \cr
                    &                               &                       \cr
Beam Momentum       & 250 GeV/c                     &   500 GeV/c           \cr
Beam Type           & $\pi^{\pm}, \ K^{\pm}, \  p$  &   $\pi^-$             \cr
Target Foil Types   & W, \ Cu, \ Al, \ Be           &   Pt, \ Diamond       \cr
No. of Target Foils & 26                            &   5                   \cr
Drift Chamber Gas   &
$49$\perc \ $Ar$, \ $49$\perc \ $C{_2}H_6$, \  $1.5$\perc \ $C_2H_5OH$ &
$89$\perc \ $Ar$, \  $10$\perc \ $CO_2$, \ $1$\perc \ $CF_4$                \cr 
Downstream $\mu^{\pm}$ Planes & 1                   &   2                   \cr
Silicon $\mu$-strip Planes & 13                     &   23                  \cr
Detector Channels   &        17000                  &   24000               \cr
Typical Event Size  &        3200 bytes             &   2400 bytes          \cr
Readout Time        &        840 $\mu$S          &   50 $\mu$S        \cr
Tape Writing Speed  &        0.5 MB/sec             &   9.6 MB/sec          \cr
Tape Change Interval &  6 minutes                   &   3 hours             \cr
Tapes Written       &    9000                       &  24000          \cr
Events to Tape &        400 million            &   20 billion          \cr
}}$$
}
2 \% of the incident pions
to interact. Thin targets also provide a better chance for short-lived
particles like the
$\Lambda_c^+$ and $\Xi_c^0$ to decay in air.
The two different materials, platinum (A=195) and
carbon (A=12), allow a measurement of the dependence of the charm
cross section on atomic number. 

   Four new silicon microstrip planes were added to a pair of silicon
planes and eight PWC planes already in place for beam tracking. The
motivation was to determine the transverse position of the primary vertex
in each event with little ambiguity. 
\vskip 10pt
\leftline{\bf SILICON MICROSTRIPS AND SPECTROMETER}
\par
   For E791 six silicon microstrip planes have been added to our
downstream SMD [6] bringing the downstream plane total to 17.
This added tracking redundancy should increase the efficiency for
reconstructing charm particles. The rest of the tracking system consists
of two analysis magnets, two PWC planes, and 35 drift chamber planes spread
among 4 DC modules.
The only new addition to the drift chambers is the non-flammable gas
shown in Table 1. Two gas \v Cerenkov counters identify kaons,
pions, and protons [7]. Calorimeters [5] provide provide electron
identification and $\pi^0$ reconstruction.  A second downstream wall of
muon scintillators has been added to help associate muon hits with
tracks. 
\vskip 10pt
\leftline{\bf HIGH SPEED PARALLEL DATA ACQUISITION}

   The DA system [8] was crucial to recording a high statistics charm
sample. \, 24000 channels were digitized by ADCs, TDCs, and latches;
and
then read out by 96 parallel controllers.  Once data was buffered into a
controller, we could take another trigger. Digitization and
readout typically
took 50 $\mu$S.   The Tevatron had 23s spill and 34s
interspill periods. Events were taken at the rate of 9000 per second
during the spill. $\!$ Eight parallel 80MB FIFO memories were used to buffer the
Tevatron cycle. Each FIFO was attached to a specific set of front end
controllers and contained a particular segment of each event. The FIFOs
were read out continuously by a six crate VME system containing a total of 
54 CPU
cards.  Each VME crate was attached to each FIFO to form a $6\times 8$
switching matrix. Six events could thus be built in parallel.  
The CPUs also compressed
events and prepared them for writing to tape.  Seven 8mm Exabyte 8200 tape
drives were attached to each VME crate.  Thus, a total of 42 Exabyte tape
drives continuously recorded events in parallel at 9.6 MB/sec. Each 8mm
tape holds 2.3 gigabytes, 13 times as much as a 9-track tape at half the cost 
per tape. When 3~hours of beam time elapsed, all 42 tapes were changed at once. 
\par
   E791 had a higher particle multiplicity and added 40\% more channels
than E769. We nevertheless made the event size 25\% smaller than E769 by
suppressing ADC zeroes and reducing the SMD word size from 16 to 
8-bits. $\!$ TDC
and ADC word sizes were held at 16- bits by using Phillips 10C6 TDCs and
LeCroy FERA 4300B ADCs.  Smaller event records increased the charm density
on tape. 
\par
   During a six month run from July 1991 until January 1992, 20 billion
physics events were recorded on 24000 tapes.  We should be able to
substantially exceed our goal of a minimum of 100,000 fully
reconstructed charm particles. 
\vskip 10pt
\leftline{\bf RISC FILTERING, CHARM PEAKS, AND PHYSICS POTENTIAL}
\par
   The Ohio State University and the University of Mississippi are each
running 1000 MIP farms composed of stripped down DECstation 5000 Model
200 workstations.  Each workstation contains a 25 MHz MIPS R3000 RISC
CPU. The hardware architecture of these farms is similar to the system
of RISC CPUs that was used previously by E769 [9]. The Mississippi
DECstations can be seen in operation in Fig. 2.  Expansions are planned
at both universities. Farms to be built at CBPF--Rio de Janeiro and Fermilab
are planned. About a third of the computing power necessary to
analyze the experiment in a timely fashion is running now. 

    We have some preliminary results, two months after the end of
our run.  A $K^0\sss \rightarrow \pi^+\pi^-$ mass plot appears in Fig.
3. \, Fig. 4 shows a mass plot of the decay $D^0 \rightarrow K^-\pi^+$.
Fig. 5 has two $D^+ \rightarrow K^-\pi^+\pi^+$ mass plots. Charge
conjugate states are implicitly included. The lower $D^+$ plot only uses
central drift chamber tracks which pass through both analysis magnets.
A significant separation, \thinspace ${\sigma}(\Delta{z})$, 
between primary and secondary vertices and $p_t$ balance
around the decaying particle direction is required in
all mass plots as shown. All plots use a 
preliminary tracking program and a rough, single bend point
approximation for magnetic fields. \v Cerenkov particle ID information
was not used, but will be later. 
\par
   When analysis
proceeds, we intend to explore a number of physics topics such as:
\begin{list}{$\bullet$}{
\setlength{\leftmargin}{0.7cm} 
\setlength{\itemsep}{0pt}
\setlength{\topsep}{1pt}
}
\item{Studying and measuring the lifetimes of charm [10] and charm--strange 
       [11] baryons in decays such as $\Lambda_c^+ \rightarrow 
       pK^-\pi^+$, \ $\Lambda_c^+ \rightarrow 
       pK^0\sss$, \ $\Lambda_c^+\rightarrow \Lambda \pi^+$, \
       and  $\Xi^0_c \rightarrow \Xi^-\pi^+$.}
\item{Finding branching ratios for doubly Cabibbo suppressed decays [12] 
       like $D^+ \rightarrow K^+\pi^+\pi^-$ and $D^0 \rightarrow K^+\pi^-$
       and for singly suppressed decays like $D^0 \rightarrow
       K^0\sss K^0\sss$ and $D^+\sss \rightarrow \phi{K^+}$.} 
\item{Measuring Cabibbo suppressed semi-leptonic decays to find the
       V(cd)/V(cs) KM matrix ratio (e.g. BR($D^0 \rightarrow
     \pi^-\ell^+\nu$) / BR($D^0 \rightarrow K^-\ell^+\nu$) )
     [13,14].}
\item{Searching for $D^0\Leftrightarrow\overline{D}{\thinspace}^0$ mixing, 
       using the $\pi$ in 
       $D^{*+} \rightarrow \pi^+D^0$ to tag the $D^0$ [15].}
\item{Searching for flavor changing neutral currents in decays such as
       $D^+ \rightarrow \pi^+\mu^+\mu^-$.}
\item{Searching for CP violation in decays such as $D^0 \rightarrow
       K^+K^-$ [16].}
\end{list}
\vskip 5pt \noindent
  This work was supported by the
U.S.\ DOE and NSF, 
the U.S.--Israel Binational Science Foundation,
and the Brasilian Conselho Nacional de Desenvolvimento Cient{\'i}fico e
Tecnol{\'o}gico.
\vskip 10pt
\leftline{\bf REFERENCES}
%
\def\issue(#1,#2,#3){\space$\underline{#1}$\space(#2)\space#3}
\def\PRL(#1,#2,#3){ Phys. Rev. Lett.\issue(#1,#2,#3)}
\def\PL(#1,#2,#3){ Physics Letters B\issue(#1,#2,#3)}
\def\PLB(#1,#2,#3){ Physics Letters\issue(#1,#2,#3)}
\def\PR(#1,#2,#3){ Phys. Rev.\issue(#1,#2,#3)}
\def\NC(#1,#2,#3){ Lettere Al Nuovo Cimento\issue(#1,#2,#3)}
\def\NIM(#1,#2,#3){ Nucl. Inst. Meth.\issue(#1,#2,#3)}
\def\NP(#1,#2,#3){ Nuclear Physics\issue(#1,#2,#3)}
\def\ZP(#1,#2,#3){ Zeitschrift f\"ur Physik C\issue(#1,#2,#3)}
\def\IEEE(#1,#2,#3){IEEE Trans. Nucl. Sci.\issue(#1,#2,#3)}
\newcounter{bean}
\begin{list}
{[\arabic{bean}]}{\usecounter{bean} \setlength{\leftmargin}{6.0mm}
\setlength{\itemsep}{0pt}
\setlength{\topsep}{1pt}
}
\item {B.H.\ Denby et al., \, Inelastic and Elastic
        Photoproduction of J/$\psi$(3097), \PRL(52,1984,795).}
\item
{J.R. Raab et al., Measurement of the $D^+$, $D^0$, and $D_{\sss}^{\lp}$
     Lifetimes, \PR(D37,1988,2391).}
\item {R.\ Jedicke and L.\ Lueking, 
     Charm Production in 250 GeV/c Hadron-Nucleon Interactions,
     Proceedings
     of the Vancouver Meeting of the DPF $\underline{2}$ (17-22 Aug 1991) 690.}
\item {R.K. Ellis and C. Quigg, A Pinacoteca of Cross-Sections 
for Hadroproduction of Heavy Quarks, FERMILAB-FN-445 (Jan 1987).}
\item {V.K.~Bharadwaj et al.,\NIM(155,1978,411);\quad \,
     V.K.~Bharadwaj et al.,\NIM(228,1985,283);\quad
     D.J.~Summers,\NIM(228,1985,290);         \quad
     J.A.~Appel et al., \NIM(A243,1986,361).}
\item {P. Karchin et al., \IEEE(32,1985,612).}
\item {D.~Bartlett et al., \NIM(A260,1987,55).}
\item
{S.~Amato, J.R.T.~de Mello Neto, J.~de Miranda, C.~James, 
\hbox{D.J.~Summers, and~S.B.}
Bracker, $\!$ The E791 Parallel Architecture Data Acquisition System, $\!$
\hbox{submitted to NIM A.}}
\item
{C. Stoughton and D. J. Summers,  Using Multiple RISC CPUs in Parallel
      to study Charm Quarks, Computers in Physics, accepted for publication.}
\item {J.C.\ Anjos et al.,
        Measurement of the $\Lambda_c^+$ Lifetime,
        \PRL(60,1988,1379).}
\item {D. MacFarlane of ARGUS reported $11.5 \pm 4.3$ \  $\Omega^0_c 
      \thinspace \rightarrow \Xi^-K^-\pi^+\pi^+$ ssc baryons at this
      conference.}
\item {J.C. Anjos et al.,
     Study of the Doubly Cabibbo Suppressed Decay $D^+ \rightarrow \phi K^+$
     and the Singly Cabibbo Suppressed Decay $D_{\sss}^{\lp} \rightarrow \phi
     K^+$, FERMILAB-PUB-91-331 (Nov 1991) submitted to Phys. Rev. Lett.}
\item {Steve Culy (Fermilab E687), Observation of the Semileptonic Decay
     $D^+ \rightarrow \overline{K}^{*0}\mu^+\nu{_\mu}$, Proceedings of the
     Rice Meeting of the DPF, Houston (3--6 Jan 1990) 546.}
\item {J.C.\ $\!$ Anjos $\!$ et $\!$ al., $\!$
        \hbox{Measurement of the Form-Factors in the Decay, $\!$
 $D^+ \! \! \rightarrow \! \overline{K}^{*0} e^+ \nu_e$, $\!$ Phys.} 
 Rev.~Lett. \issue(65,1990,2630); \quad J.C. Anjos, Study of the Decay $D^+ 
        \rightarrow \overline{K}^{0} e^+ \nu_e$, \PRL(67,1991,1507).}
\item {J.C.\ Anjos et al.,
        A Study of $D^0$--$\overline{D^0}$ Mixing from Fermilab E--691,
        \PRL(60,1988,1239); \quad
        L. Wolfenstein, \PLB(164B,1985,170); \quad
        A. Datta, \PLB(154B,1985,287); \quad
        J.F. Donoghue et al.,\PR(D33,1986,179).}
\item {J.C. Anjos et al.,
     Measurement of the decays $D^0 \rightarrow \pi^-\pi^+$ and 
     $D^0 \rightarrow K^-K^+$, \PR(D44,1991,R3371).}
\end{list}
\vspace*{6mm}
\leftline{\small \hspace{48mm} 
Pt \hspace{4mm} C \hspace{5mm} C \hspace{5mm} C \hspace{5.5mm} C}
\vspace*{-9mm}
\leftline{\hspace{27mm}
\resizebox{95mm}{!}{\includegraphics*[25,55][530,530]{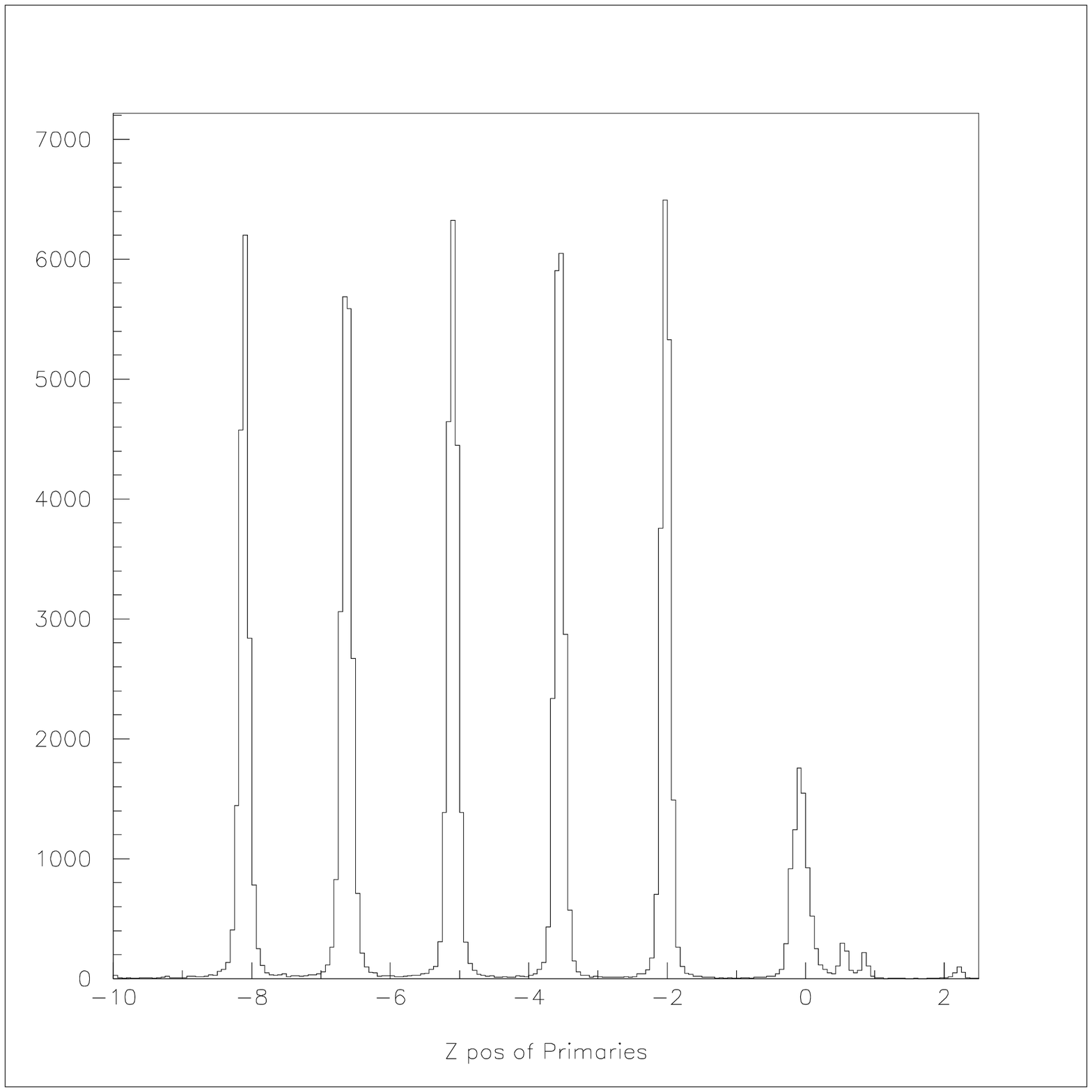}}}
\vspace*{-1mm}
\centerline{\bf \hspace{1mm} Centimeters}
\vspace*{1mm}
\centerline{{\bf{Fig. 1.}} {\small Primary vertices from tracks show the
{\it z} positions of our targets, 1 Pt and 4 diamond.}} 
\vskip 4pt
\vspace*{6mm}
\leftline{\hspace{21mm}
\resizebox{105mm}{!}{\includegraphics{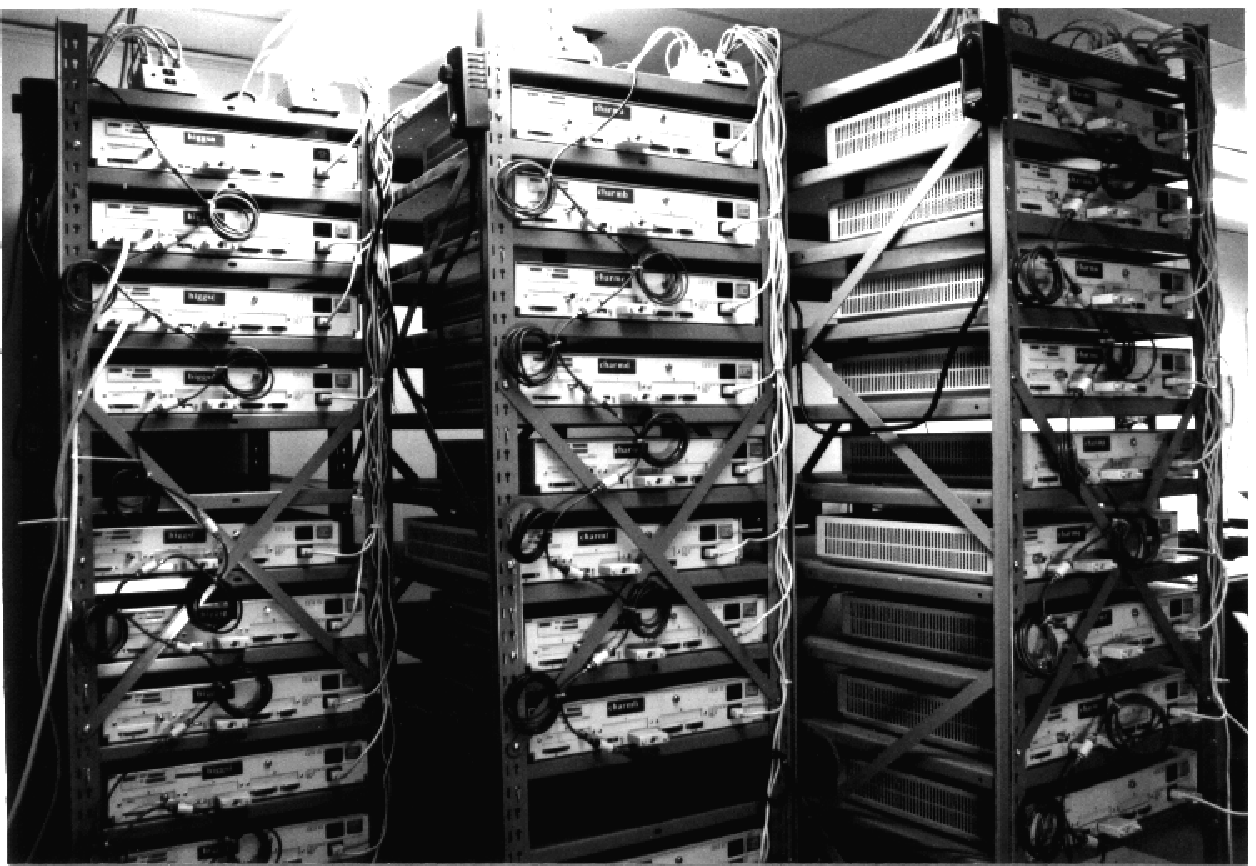}}}
\vfill
\vbox{\noindent{\bf Fig. 2.} Rack mounted workstations 
filtering data; a very cost effective means of computing. A pair of 10
tape Exabyte EXB-10  
\hbox{autoloaders provides an input buffer of over 40 gigabytes.}
Ethernet distributes events.  
\hbox{These particular workstations are DECstation 5000/200s with}
{\small{MIPS R3000 CPU}}s. Other current possibilities include the 
\hbox{Hewlett Packard-Apollo~9000/705,}
the DECstation 5000/25, the Silicon Graphics Indigo,
the IBM RS/6000--320H, and the Sun SPARCstation 2. Running your own benchmarks
and comparison shopping is often useful.} 
\eject
\vbox{
\vspace*{13mm}
\leftline{\LARGE \hspace{8mm} $K^0_S$}
\vspace*{-12mm}
\leftline{\hspace{-2mm}
\resizebox{68.1mm}{!}{\includegraphics*[25,150][550,660]{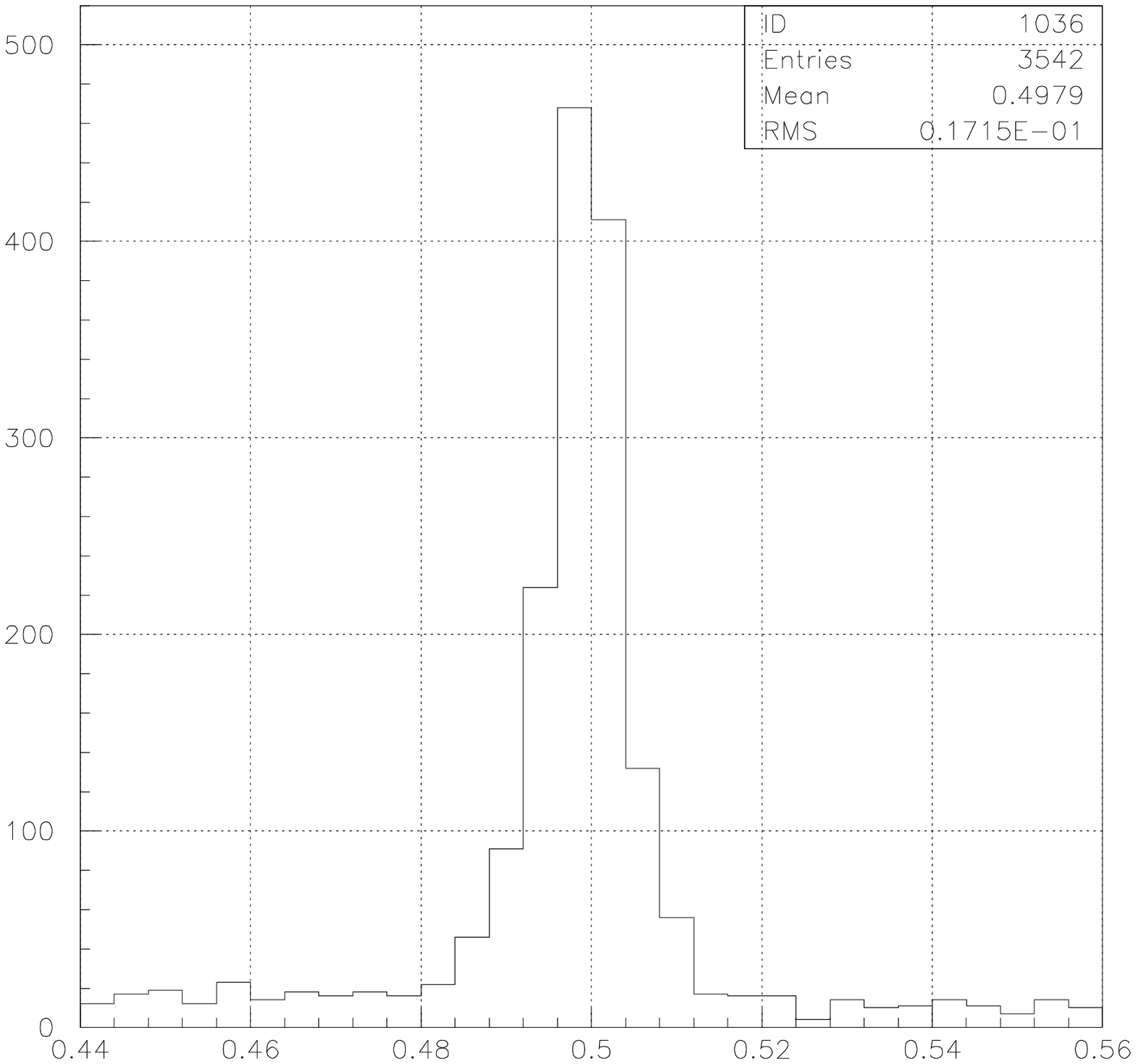}}}
\vspace*{0mm}
  \vbox{
  \leftline{\qquad \qquad \bf Mass($\pi^+\pi^-$) GeV/c$^2$}
  \vskip 8pt
  \leftline{{\bf{Fig. 3.}} $K^0\sss{(498)} \rightarrow \pi^+\pi^-$ mass plot.}
  \leftline{Cuts: ${\sigma}(\Delta{z}) > 15$ \quad \quad \quad 
  cos$\thinspace \theta < 0.8$}
  \leftline{$p_t$ unbalance around $K^0\sss < 0.1$ GeV/c.}} \break
\vspace*{-63mm}
\leftline{\hspace{77.5mm}
\resizebox{75mm}{!}{\includegraphics*[40,300][280,525]{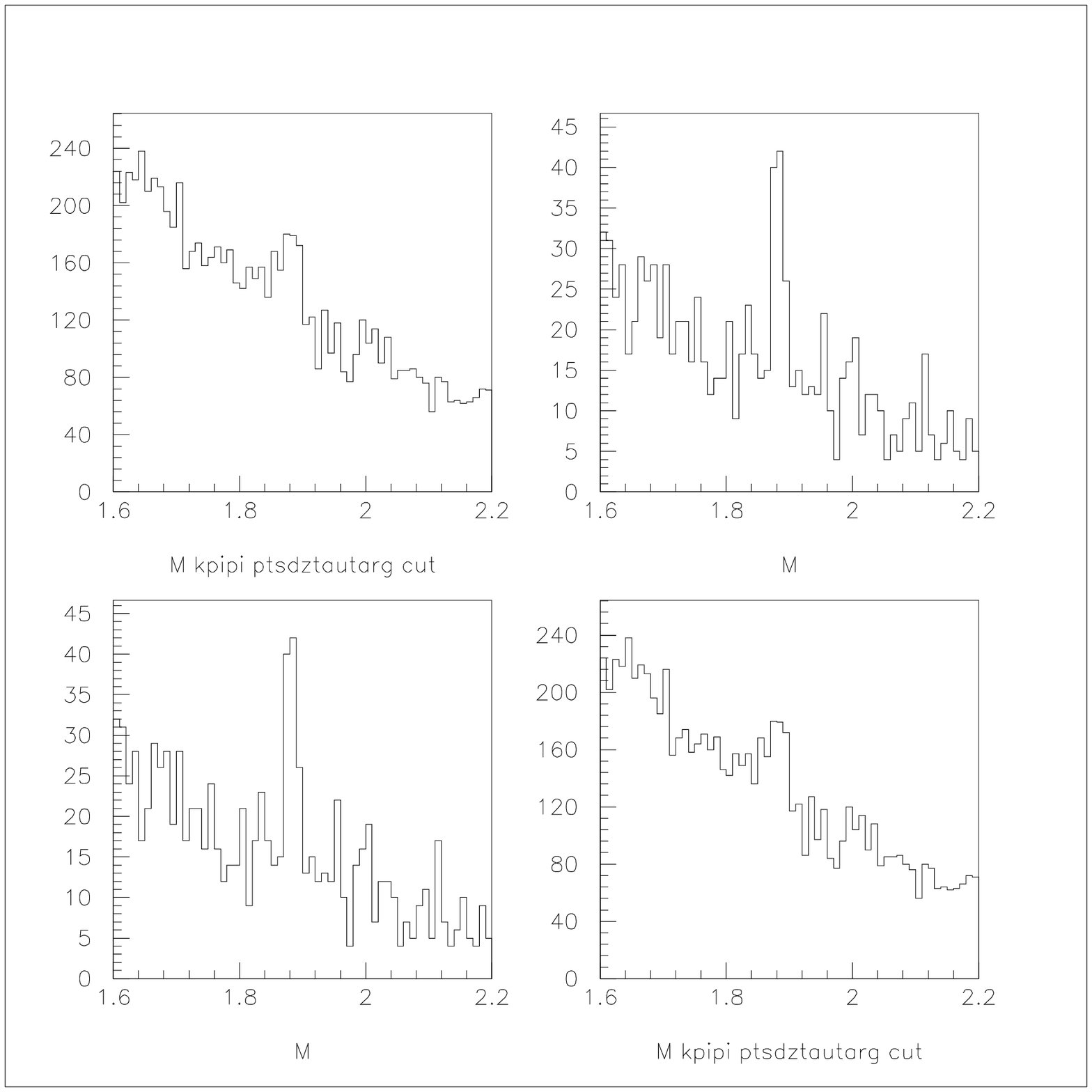}}} 
\vspace*{5mm}
\leftline{\hspace{-1mm}
\resizebox{75mm}{!}{\includegraphics*[290,300][530,530]{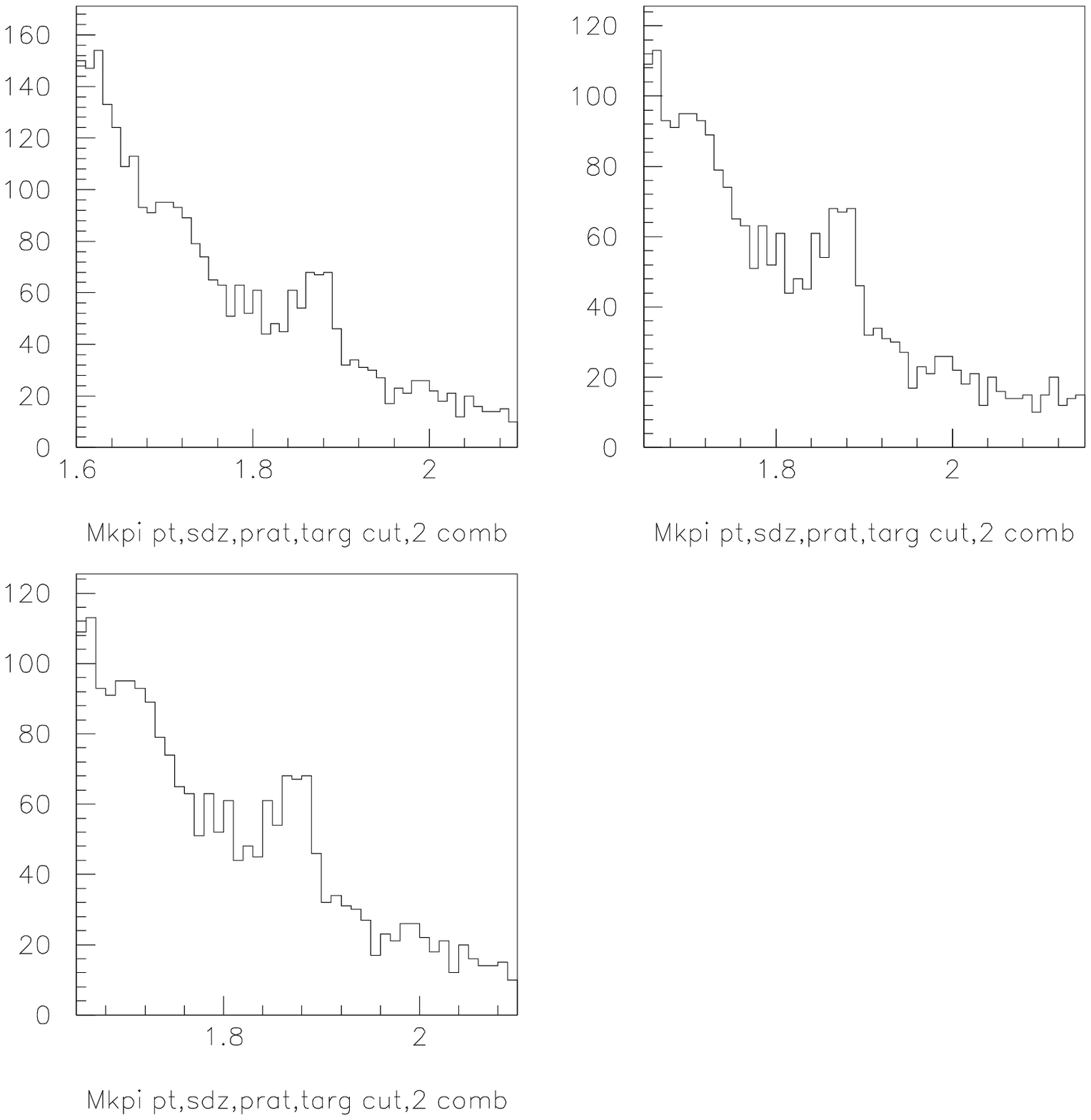}}
\hspace{1mm}
\resizebox{75mm}{!}{\includegraphics*[40,50][280,275]{dplus.ps}}}
\vspace*{-10mm}
$$\vbox {\halign{#\hfil & \ #\hfil \cr
\qquad \qquad \bf Mass($K^-\pi^+$) GeV/c$^2$                         &
\qquad \qquad \bf Mass($K^-\pi^+\pi^+$) GeV/c$^2$                    \cr
                                                             &       \cr
{\bf Fig. 4.} $D^0(1865) \rightarrow K^-\pi^+$ mass plot            &
{\bf Fig. 5.} $D^+(1869) \rightarrow K^-\pi^+\pi^+$ mass plots    \cr 
Cuts: ${\sigma}(\Delta{z}) > 10 \quad \quad \quad \tau < 1.6$ ps     &
Cuts: ${\sigma}(\Delta{z}) >  8 \quad \quad \quad \tau < 3.0$ ps     \cr
$p_t$ unbalance around $D^0 <  0.35$ GeV/c.                           &
$p_t$ unbalance around $D^+ <  0.35$ GeV/c.                           \cr
Distance to any target center $>$ 1.0 mm.                             &
Distance to any target center $>$ 1.0 mm.                             \cr
Secondary vertex distance of closest approach                        &
The bottom plot alone demands hits in all                            \cr
/ primary vertex DCA $<$ 0.75 for both tracks.                       &
4 drift chamber modules for all 3 tracks.                            \cr
}}$$
}
%
%
%
%
%
\end{document}